\documentclass[aps,prb,preprint,nopacs,superscriptaddress,floatfix]{revtex4-1}
\usepackage{graphicx}
\usepackage{verbatim}
\usepackage{amsmath}
\usepackage{sidecap}
\usepackage{epstopdf}
\usepackage{bm}

\usepackage{amsmath, amssymb, graphics, mathrsfs, bm, stackrel,bbold}

\usepackage[usenames,dvipsnames]{xcolor}
\definecolor{Highlight}{rgb}{1,1,0.75}
\definecolor{Maintext}{rgb}{0.05,0.05,0.05}

\usepackage[export]{adjustbox}

\newcommand\ba{\begin{array}}
\newcommand\ea{\end{array}}
\newcommand\nn{\nonumber}
\newcommand\ri{\right}
\renewcommand\le{\left}

\newcommand{\feyn}[1]{#1\kern-0.45em/}

\newcommand{\tto}{\rightarrow}
\newcommand{\sgn}{\mbox{sgn}}

\renewcommand\d{\delta}

\newcommand\e{\epsilon}

\renewcommand\L{\Lambda}

\newcommand\p{\pi}

\newcommand\s{\sigma}

\newcommand\la{\langle}
\newcommand\ra{\rangle}

\newcommand\mc{\mathcal}

\begin{document}

\title{Connection topology of step edge state bands at the surface of a three dimensional topological insulator}

\author{Yishuai Xu}
\affiliation{Department of Physics, New York University, New York, New York 10003, USA}
\author{Guodong Jiang}
\affiliation{Department of Physics and Astronomy, Purdue University, West Lafayette, IN 47907, USA}
\author{Janet Chiu}
\affiliation{Department of Physics, New York University, New York, New York 10003, USA}
\author{Lin Miao}
\affiliation{Department of Physics, New York University, New York, New York 10003, USA}
\affiliation{Advanced Light Source, Lawrence Berkeley National Laboratory, Berkeley, CA 94720, USA}
\author{Erica Kotta}
\author{Yutan Zhang}
\affiliation{Department of Physics, New York University, New York, New York 10003, USA}
\author{Rudro R. Biswas}
\email{rrbiswas@purdue.edu}
\thanks{Corresponding author}
\affiliation{Department of Physics and Astronomy, Purdue University, West Lafayette, IN 47907, USA}
\author{L. Andrew Wray}
\email{lawray@nyu.edu}
\thanks{Corresponding author}
\affiliation{Department of Physics, New York University, New York, New York 10003, USA}

\begin{abstract}

Topological insulators in the Bi$_2$Se$_3$ family manifest helical Dirac surface states that span the topologically ordered bulk band gap. Recent scanning tunneling microscopy measurements have discovered additional states in the bulk band gap of Bi$_2$Se$_3$ and Bi$_2$Te$_3$, localized at one dimensional step edges. Here numerical simulations of a topological insulator surface are used to explore the phenomenology of edge state formation at the single-quintuple-layer step defects found ubiquitously on these materials. The modeled one dimensional edge states are found to exhibit a stable topological connection to the two dimensional surface state Dirac point.

\end{abstract}


\date{\today}

\maketitle

\section{Introduction}\label{sec:introduction}

Three-dimensional topological insulators (TI) are materials with \textbf{Z$_2$} topological order that manifest conducting two-dimensional (2D) Dirac cone surface states protected by time reversal symmetry \cite{Zahidreview,FuOriginal}. The surface state electrons resist scattering from weak non-magnetic perturbations \cite{Zahidreview,SCZhangreview,FuOriginal,HsiehScience2009,roushanNature2009,MoorePRB2007,roushanNature2009,AlpichshevPRL2010,QiPRB2008}, however recent studies have shown that strongly perturbing point- and step- like surface defects can introduce new in-gap states and modify the band structure near the Dirac point \cite{ShafferPRB2014,BiswasPRB2010,TeagueSolidState2012,AlpichshevPRL2012,SchafferPRB2012,SchafferPRB2012_2,AlpichshevPRB2011,DmitrievJETP2014,WangPRB2011,BiswasPRB2011,RizzoPRB2013,FuPRB2011,fu2014anisotropic,LiuPRB2012,Xu2017disorder}. Here, we present a numerical and analytic analysis of the single quintuple-layer step defect of Bi$_2$Se$_3$-family TIs, to explore the nature of the associated edge state. These investigations establish that edge states with a stable topological connection to the 2D Dirac point of the TI surface state can exist, in scenarios that do not require fine tuning of the Hamiltonian. As with the surface states of Weyl semimetals, the connection is defined by a linear dispersion of the lower-dimensional (here 1D) state, converging on the Dirac point of higher dimensional (2D) band structure. The protected nature of this connection is considered with respect to broken symmetries and disorder, and it is shown that the occurrence of such states can be ubiquitous across a wide range of parameters for describing surface steps in a 3D TI tight binding model. 

\begin{figure}[t]
\includegraphics[width = 7cm]{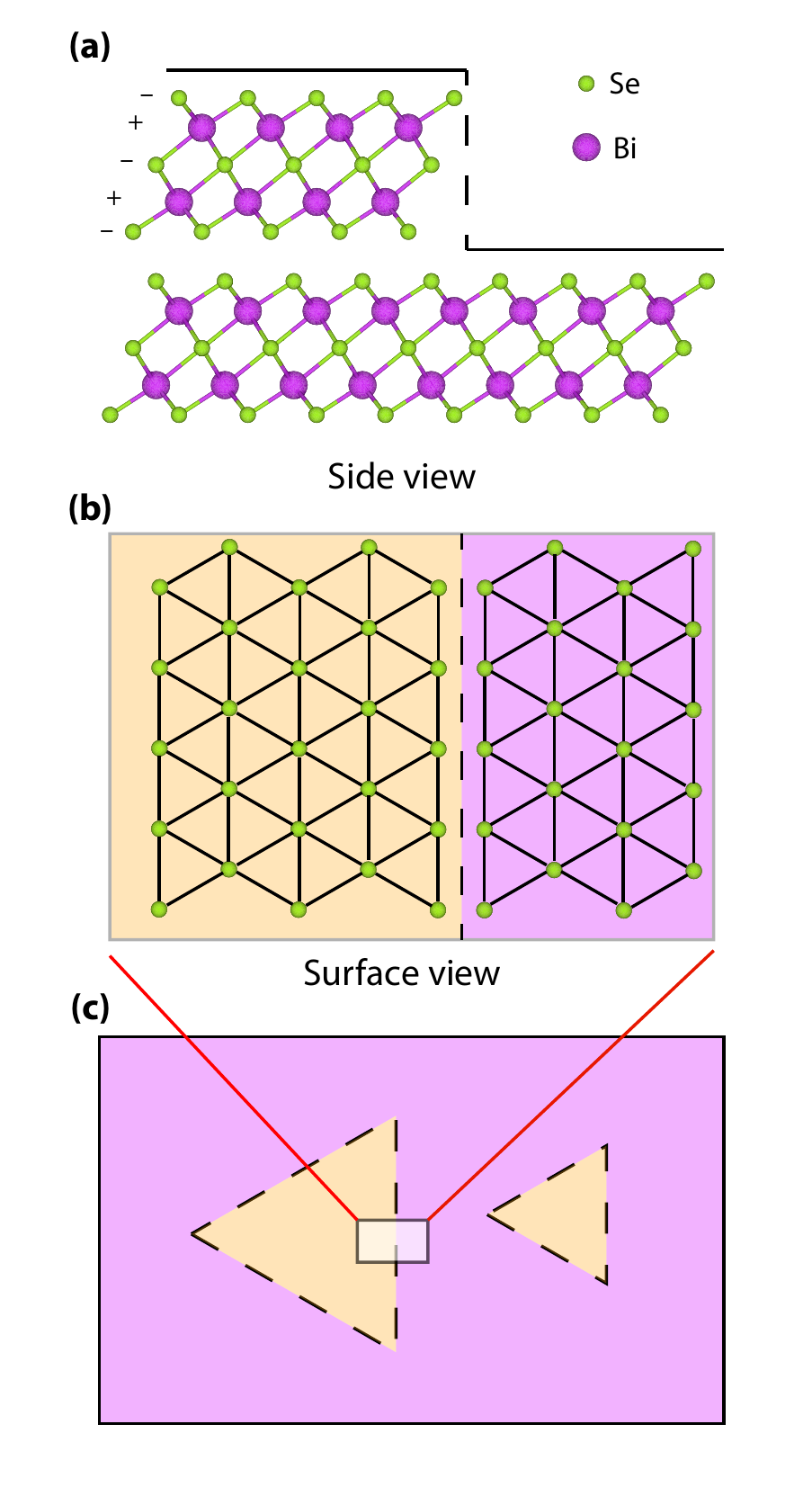}
\caption{{\bf{Structure of a Bi$_{2}$Se$_{3}$ surface step}}: (a) Side view of a single quintuple layer step parallel to the in-plane nearest neighbor axis. (b) Sample surface showing outermost Se atoms near a single quintuple layer step. (c) Large scale schematic of the triangular step plateaus found on Bi$_2$Se$_3$. The image is not drawn to scale, as real terraces tend to be over 100 nm in size.}
\label{fig:Structure}
\end{figure}

Bi$_{2}$Se$_{3}$ is widely seen as a model system for studying TI surface physics, as it has one of the largest band gaps presently known in a TI system ($\sim 300$ meV), which is spanned by a relatively ideal single Dirac cone surface state \cite{ZhangNatphys2009,XiaNatphys2009,ChenScience2009}. The crystal structure of Bi$_{2}$Se$_{3}$ is shown in Fig. \ref{fig:Structure}(a), with weak Van der Waals bonding between stacked quintuple atomic layers. Step defects one quintuple layer in height are very common at the surface of thin film samples in this material as shown in Fig. \ref{fig:Structure}(a) \cite{TeagueSolidState2012,HarrisonAPL2013}. Fig. \ref{fig:Structure}(b) shows the arrangement of Se atoms at a cleaved surface near a single step, seen from above. Viewed on a micron scale, the steps tend to run parallel to the in-plane nearest-neighbor axis, resulting in triangular plateaus as shown in Fig. \ref{fig:Structure}(c) \cite{TeagueSolidState2012,HarrisonAPL2013}. 

Protected 1D edge states are typically associated with 2D topological order, however several classes have also been proposed to occur at the surfaces of 3D material systems under specific circumstances. 
Hinge states can occur at the intersection of two non-parallel faces of a so-called `higher order topological insulators' \cite{BernevigEdgeTI, HOTI}, and are expected in special cases for traditional 3D TIs \cite{TIsurface_junction}. In the 3D TI case, if the intersecting faces have Dirac points at the same energy, the edge can host an in-gap mode that intersects the 2D Dirac point, associated with a near-realization of the Jackiw-Rebbi Hamiltonian \cite{Jackw_Rebbi}. However, structurally simple intersections of this type are challenging to create and study experimentally. Another topologically protected edge state scenario has been identified at certain classes of step edges on a topological crystalline insulator (TCI) surface, and can be understood from extrapolation to a scenario in which particle-hole symmetry is unbroken \cite{sessi2016robust}.

Other 1D in-gap bound states that converge on the Dirac point of a massless 2D Dirac Hamiltonian have been identified for specific models, such as the bound states underneath a gate electrode \cite{Nagaosa_gate} or a 1D Gaussian potential \cite{Gaussian_2014PRB}. These scenarios are intriguing because although they do not require fine tuning of the Hamiltonian, they nonetheless appear to be almost \emph{coincidental}, and have not been identified explicitly with a distinct topological invariant of the system. The principal result of this paper will be to show that this class of topologically connected edge states is insensitive to the specific form of the TI Hamiltonian or the 1D feature to which edge states are bound, and is expected to be a generic feature of line-like defects on the surface of 3D TIs (or, equivalently, planar defects in a 3D Weyl kinetic Hamiltonian).

This paper is organized as follows: In Sec. \ref{sec:existence of edge state}, we establish a broad set of conditions allowing the existence of edge states with non-trivial band connectivity, bound to a 1D scalar potential. In Sec. \ref{sec:latticeModel}, we present numerical simulations on a 2D lattice model motivated by real TI surfaces, and demonstrate the robustness of the connectivity of 1D edge states against symmetry breaking. In Sec. \ref{sec:tight binding}, we show examples of this topological connectivity within a 3D tight binding model resembling Bi$_2$Se$_3$.

\section{Allowed existence of an edge state with non-trivial band connectivity}\label{sec:existence of edge state}

\begin{figure}[t]
\vspace{1.7 cm}
\includegraphics[width = 12cm]{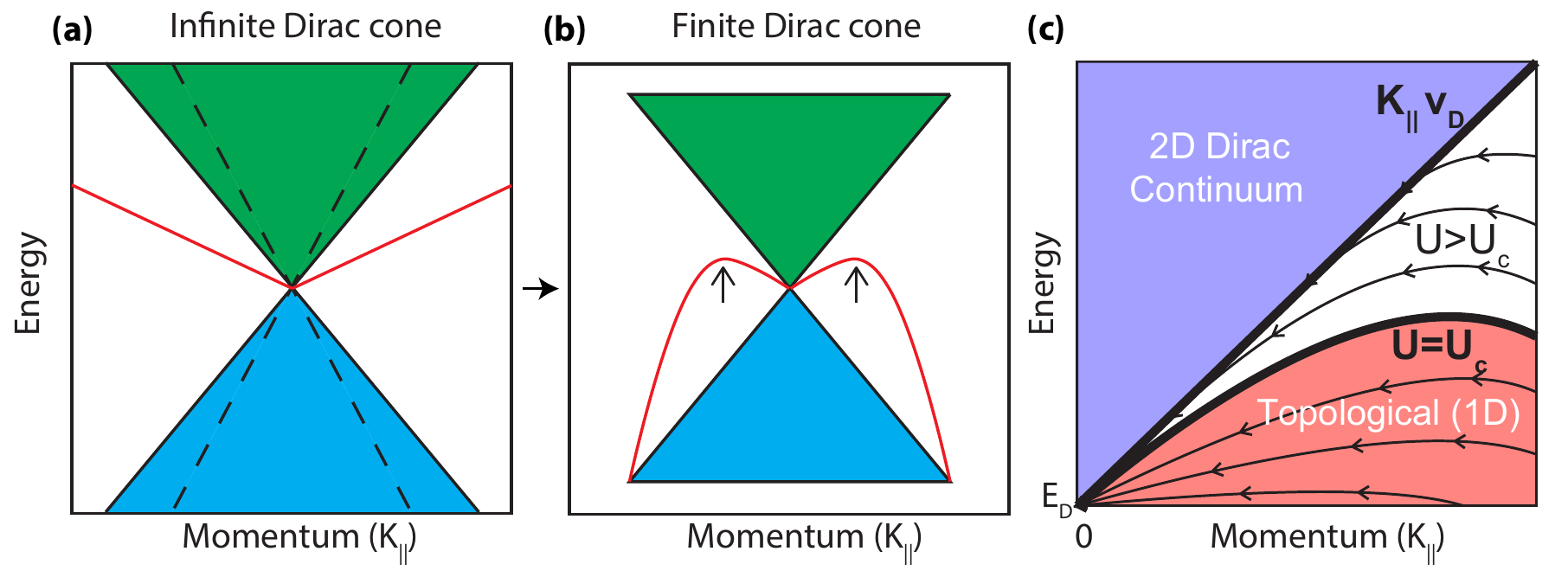}
\caption{{\bf{Topological connectivity of a 1D edge state}}: Step edge states on a surface with an (a) infinite and (b) finite 2D Dirac cone. Edge state dispersions that are non-degenerate with the 2D band continuum are traced in red and ungapped (ill-defined) edge state dispersions are shown with dashed lines. (c) A dispersion diagram showing how edge states created by different 1D scalar potentials connect back to the 2D surface state Dirac point, or merge with the 2D state continuum, as $K_\parallel$ momentum approaches zero. Arrows indicate the `flow' direction as momentum is reduced, and edge states that connect with the Dirac point are labeled as `Topological'. If these states are present, they will occur for a continuous range of step edge potentials beneath a critical value ($U<U_C$), constituting a red-shaded topologically ordered region in the phase diagram.}
\label{fig:Step edge}
\end{figure} 

Bound states at quintuple-layer step edges have been observed for the related TIs Bi$_2$Se$_3$ and Bi$_2$Te$_3$ \cite{AlpichshevPRB2011,FedotovPRB2017}, and a recent STM analysis has been interpreted to suggest that they may be a form of 1D electron gas (1DEG) brought on by an effective scalar potential at the step \cite{FedotovPRB2017}. Because of this, and for the sake of analytical tractability,
We will begin by considering the scenario of a 1D delta function potential added to a 2D system defined by a massless Dirac Hamiltonian. Momentum along the 1D potential axis is a conserved quantity ($k_\parallel$), and the 1D bound state dispersion in this scenario is already known from Ref. \onlinecite{Nagaosa_gate} to converge on the 2D Dirac point with a constant group velocity.

Changing the delta function potential strength modifies the velocity of the bound state, but not the momentum-space connectivity. The connection to the 2D Dirac point can thus be called a topological property, in that it is robust for a continuous range of 1D perturbation strengths. However, the edge state connectivity is not generically protected from Coulomb perturbations, as a surface potential in the right form could in principle cause the group velocity of the edge state to exceed the Dirac velocity, pushing the edge state dispersion into the 2D Dirac cone, making it no longer a well-defined edge state (dashed lines in Fig. \ref{fig:Step edge}(a)).

The projection of the bound state onto kinetic basis states in the 2D Dirac cone is localized, and is negligible at energies outside of a $|E|>v_\L k_\parallel$ window around the 2D Dirac point, for sufficiently large values of a constant $v_\L$ (see discussion in Appendix \ref{appendix:energy_cutoff}). We observe that this has the consequence that a non-delta function potential will generally give a very similar linear dispersion converging on the Dirac point (at $k_\parallel\sim 0$). So long as the Fourier transform of the potential ($V(q)$) is nonzero at q=0 and effectively flat on a momentum scale of $\delta q\gtrsim v_\L k_\parallel/v_D$, the potential will be indistinguishable from a delta function within the low energy basis that composes the step edge state. This scenario is played out in a practical context in Ref. \onlinecite{Gaussian_2014PRB}, which establishes the linear dispersion for a Gaussian 1D potential. Qualitatively speaking, as the amplitude of $k_\parallel$ is increased, non-linearities in the surface state dispersion will emerge when the curvature of $V(q)$ becomes significant on a momentum scale proportional to $k_\parallel$, and/or when the kinetic state basis at energies within $|E|<|k_\parallel| v_\L$ deviates from a massless 2D Dirac Hamiltonian (see turning point indicated by arrows in Fig. \ref{fig:Step edge}(b)). Corrections to the edge state energy from these factors act at lowest order in proportion to $k_\parallel$ (see Appendix \ref{appendix:energy_cutoff}), meaning that they can influence the edge state velocity, but not the point of convergence.

Deviation from a massless 2D Dirac Hamiltonian is inevitable at large momentum in a real material, and must be accounted for to understand the band structure connectivity of the \emph{other end} of the 1D edge state in momentum space. If the 2D Dirac cone Hilbert space is curtailed by a high energy cutoff such that kinetic basis states outside a momentum window $|\textbf{k}|<\L$ are disregarded, the edge state will follow an arching dispersion like that drawn in Fig. \ref{fig:Step edge}(b) (see also an analogous simulation in Fig. \ref{fig:robustness}(a)). This scenario is close to what is expected at a real step edge, since the 2D Dirac cone surface states of TIs are only well defined over a small energy range inside the bulk band gap. Once the band momentum becomes sufficiently large, the edge state of a delta function potential curves in a direction opposite to the sign of the edge potential ($U$, defined below). It is required to merge with the 2D Dirac cone at or before the momentum cutoff, due to the lack of states for the 1D potential to couple between as momentum along the edge approaches the cutoff. This connection bridged by the edge state between the 2D surface Dirac point and the state continuum immediately above or below the 2D Dirac point is a further property that can be used to classify an edge state, and will be discussed later in the context of a more realistic model with full 3D topological order (Section \ref{sec:tight binding}).

The diagram in Fig. \ref{fig:Step edge}(c) shows a summary of the topological and non-topological dispersions that can be expected for a 1D edge state bound to a scalar potential parametrized by $U$. At a critical value of $U=U_C$, the edge state will converge towards the Dirac point with a velocity $v_S$ equal to the 2D Dirac velocity ($v_S=v_D$). Increasing $U$ causes the dispersion to intersect with the 2D Dirac continuum (white region), so that the edge state does not include a well defined dispersion that intersects with the Dirac point. For a continuous range of lower potentials $U<U_C$, the edge state converges on the Dirac point (red region, labeled Topological) until velocity of convergence becomes equal to the negative Dirac velocity ($v_S=-v_D$, not shown in Fig. \ref{fig:Step edge}(c)). The allowed existence of a topological connection that is protected over a finite range of constant prefactors for essentially any spatial form of the 1D potential is robust, and is not conditional on symmetries that do not destroy the Dirac point, such as positive reflection symmetry across the step, or particle hole symmetry in the kinetic Hamiltonian. Numerical simulations in the next sections will show that Dirac-point-intersecting 1D bound state dispersions are a common, and possibly ubiquitous, feature in typical models of step edges at TI surfaces.

\section{Edge states of a 1D scalar potential in a 2D lattice model}\label{sec:latticeModel}

To relate the above picture more closely to the step edge states seen in real TI materials, we consider a 2D hexagonal real space lattice that resembles the Bi$_{2}$(Se/Te)$_{3}$ surface, following the modeling implementation in Ref. \onlinecite{Xu2017disorder}. The step edge is described as a scalar potential that repeats along a chain of surface sites extending along the crystallographic a-axis (nearest neighbor direction). The modeled system has translational symmetry along this axis, and is simulated with a large (effectively infinite) number of sites along the y-axis. The non-interacting Dirac Hamiltonian for a 2D surface electron can be written as \begin{equation}
H_{T}=v_{D} (\textbf{k} \times \bm{\sigma} )
\end{equation}
where $v_{D}$ is the Dirac velocity, $k$ is the momentum of the surface electron and $\sigma$ is the Pauli vector. The Hamiltonian for the step is written as 
\begin{equation}
H_{U}=U\sum_{\alpha} n_{\alpha}
\label{eq:Udef}
\end{equation}
where $U$ is the step potential, $\alpha$ indexes the sites intersected by the step, and $n_{\alpha}$ is the on-site electron number operator. The complete Hamiltonian for the modeled system is then $H=H_{T}+H_{U}$. The Dirac velocity is taken to be 3 eV$\cdot \AA$, the hexagonal lattice constant is 4.2 $\AA$, and energy cutoff for the kinetic basis is $v_D\L=0.4$ eV. Exact diagionalization is used to obtain the eigenstates and energies of the system.

The local density of states (LDOS) on top of the modeled step edge was shown in our previous work \cite{Xu2017disorder}, and closely matches the momentum-integrated LDOS profile of a step edge state seen by STM on Bi$_2$Te$_3$, at a step potential $U=3.8$ eV. The dependence of step edge LDOS on momentum parallel to the step ($k_{\parallel}$) is shown in Fig. \ref{fig:robustness}(a), and closely resembles the qualitative expectation depicted in Fig. \ref{fig:Step edge}(b), with identical connections to the 2D Dirac continuum at the $k_{\parallel}\rightarrow0^+$ and $k_{\parallel}\rightarrow \L$ limits. Local maxima of the dispersion result in LDOS maxima (square root anomalies) approximately 90 meV above the Dirac point.

Increasing the step potential causes the edge state to converge towards the upper Dirac cone dispersion (Fig. \ref{fig:robustness}(b-c)). State dispersions closely resemble the continuum-limit (CL) expectation for a delta function potential, derived in Appendix \ref{appendix:energy_cutoff}). In this idealized case, the edge state velocity approaches the Dirac velocity as $U\rightarrow\infty$, meaning that there is no finite value for $U_C$, and the system has topological connectivity for all values of $U$. The CL model includes just one surface state, and large deviations from the model are expected to occur when a new surface state appears, as will be shown in the 3D tight binding model simulations in Section \ref{sec:tight binding}.

\subsection{Protection of the edge state}\label{sec:protection of edge state}

\begin{figure}[t]
\includegraphics[width = 12cm]{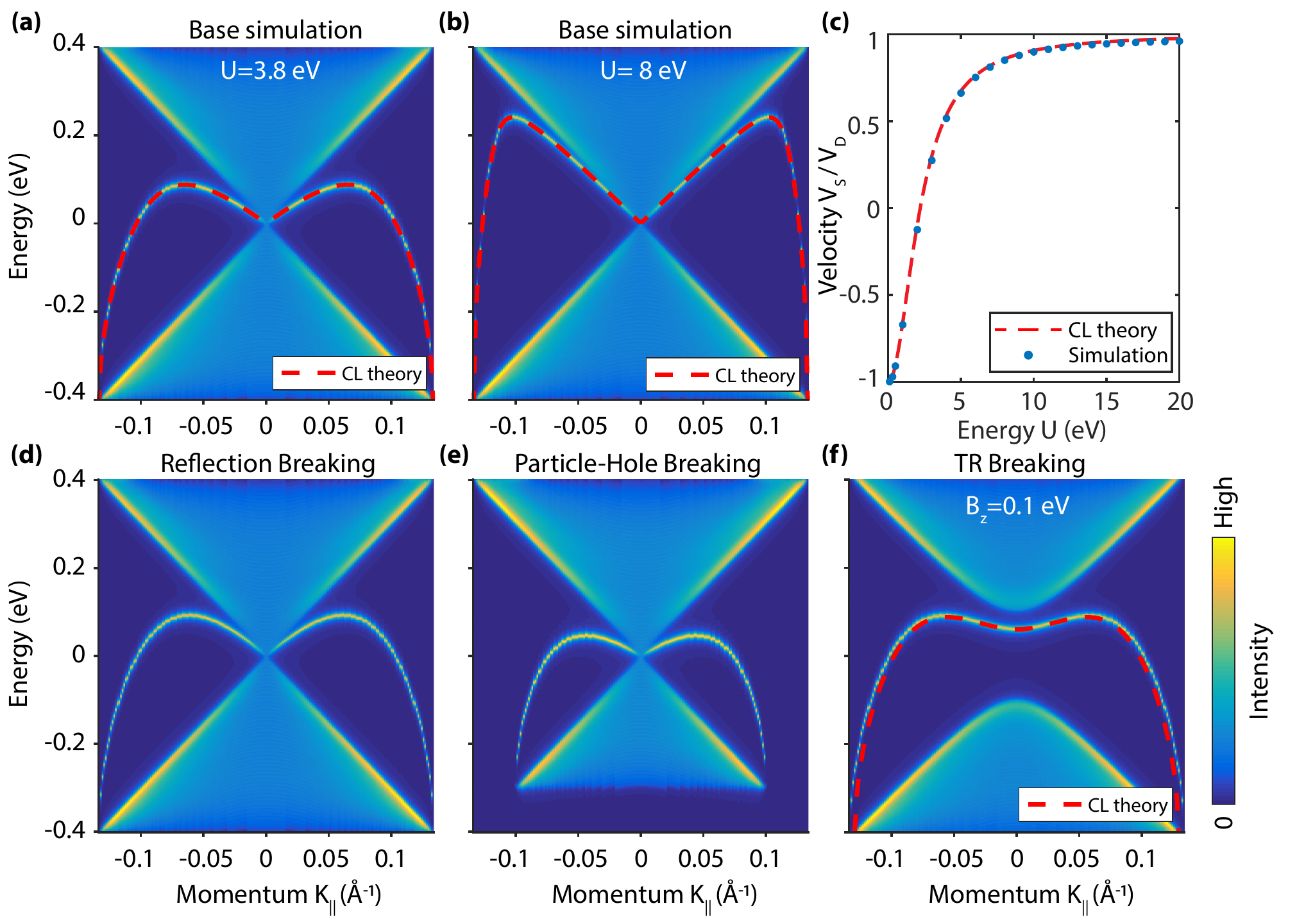}
\caption{{\bf{Robustness of the protected connectivity}}: (a) The $k_\parallel$ resolved LDOS within 21 $nm$ of a scalar-potential step edge in the discrete 2D lattice model described in Section \ref{sec:latticeModel}. Dashed lines represent results from the analytic continuum-limit (CL) theory (see Appendix \ref{appendix:energy_cutoff} and Eq. \ref{eq-gapless-bound}). The potential used to describe the step edge is $U=3.8$ eV, as fitted to STM on Bi$_2$Te$_3$ in our previous work \cite{Xu2017disorder}. (b) A similar simulation with $U=8$ eV. (c) Edge state velocity ($V_S$) at momenta slightly displaced from the 2D Dirac point ($K_\parallel \sim 0^+$) is shown as a function of the step potential ($U$) in the discrete lattice simulation and the analytic 2D continuum-limit theory. (d-f) LDOS showing step edge state dispersion for systems with $U=3.8$ eV and broken reflection, particle-hole, or time-reversal symmetry, respectively.}
\label{fig:robustness}
\end{figure}

The allowed existence of edge states with a protected connection to the Dirac point as established in Section \ref{sec:existence of edge state} is not conditional on Hamiltonian symmetries such as reflection symmetry or particle hole symmetry, so long as the 2D Dirac point itself is not destroyed, and the Fourier transform of the 1D potential does not vanish at $q=0$. For example, reflection symmetry of the 1D potential is broken in Fig. \ref{fig:robustness}(d) by parallel lines with $U=0.1$ eV and $U=-0.1$ eV one lattice site above and below the $U=3.8$ eV `step', respectively. Similarly, the simulation in Fig. \ref{fig:robustness}(e) breaks particle-hole symmetry of the kinetic Hamiltonian by reducing the negative energy cutoff to -0.3 eV. In each case, the Dirac point connectivity of the edge state is unchanged, but the velocity with which it intersects the Dirac point is slightly altered.

However when the 2D Dirac point itself is gapped by the addition of a Zeeman Hamiltonian term that breaks time reversal symmetry ($H_B=B_z\sigma_z$), the protected connection is also necessarily broken as shown in Fig. \ref{fig:robustness}(f). The edge state disperses through the gap with a new extremum at zero momentum that would create an additional square root anomaly in sufficiently high-resolution LDOS measurements. The precise dispersion can be extrapolated by noting that the Zeeman and $k_\parallel$ terms in the model Hamiltonian combine to create a Pauli vector that is orthogonal to the only other Pauli vector in the Hamiltonian (which comes from $k_\perp$). As such, the edge state energy at momentum $k_\parallel$ in the presence of a perturbing magnetic field will be identical to the unperturbed energy at a different momentum $k_\parallel'$, where $k_\parallel'=\sqrt{B_z^2+k_\parallel^2}$.

Disorder is another phenomenon that can render topological band features ill-defined, by causing momentum to no longer be a good quantum number. However, because the edge states appear in spin-chiral time reversal pairs, disorder that does not violate time reversal symmetry will only mix the edge states with bulk states, and not with the time-reversed partner. The effect of time-reversal invariant disorder is therefor expected to be proportional to the bulk density of states, which vanishes at the Dirac point. While the states may still be destroyed due to strong disorder, the connection to the Dirac point has qualitatively more protection than dispersions at other energies, suggesting that it will be a robust phenomenon.

\section{Physical steps in a 3D tight binding model}\label{sec:tight binding}

\begin{figure}[t]
\includegraphics[width = 8.6cm]{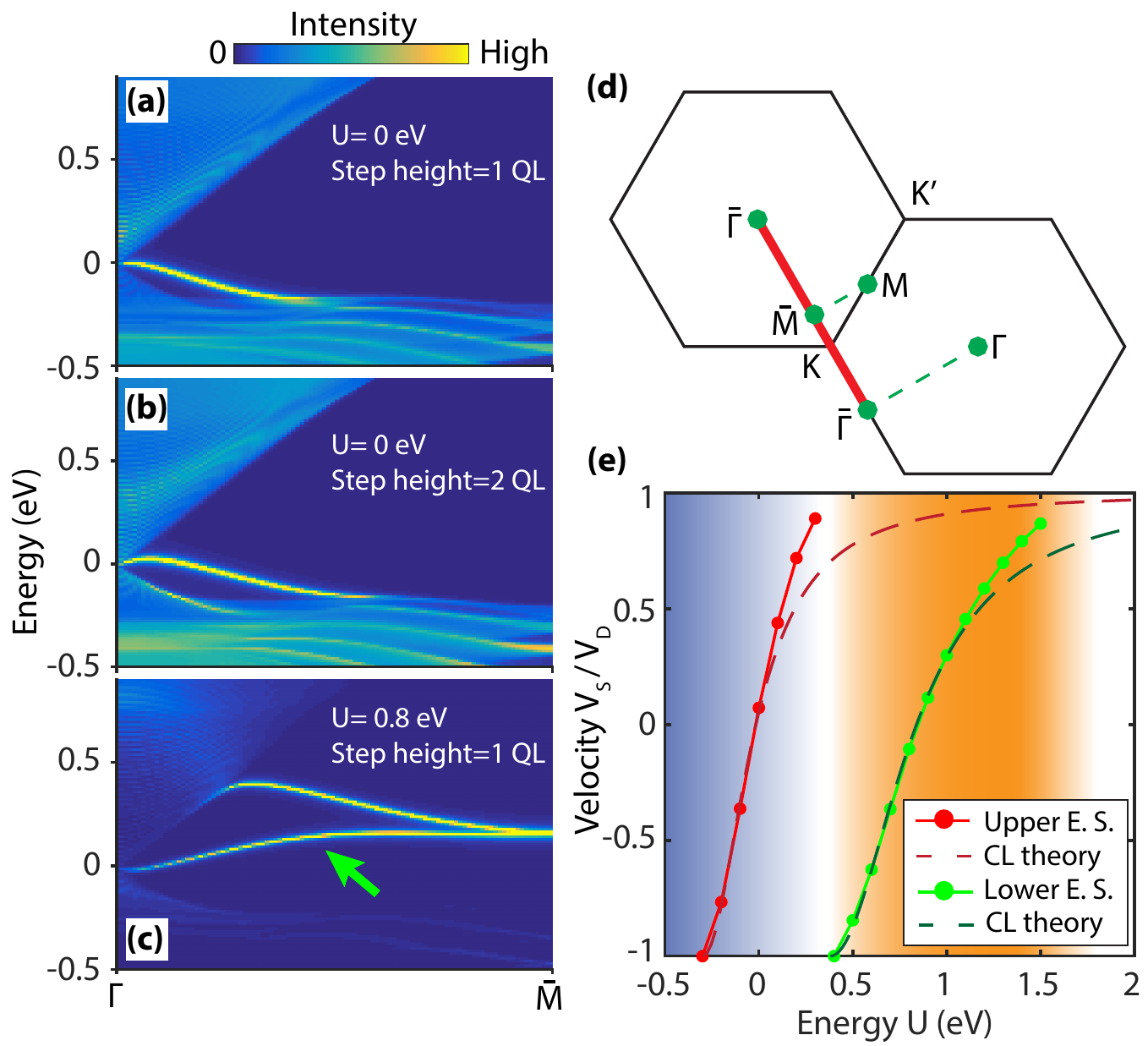}
\caption{{\bf{Step edge state in a 3D tight-binding model}}: (a-b) The $K_\parallel$-resolved LDOS at the step edge in a tight binding model with 3D TI order and a 1QL or 2QL step, respectively. No scalar potential is applied in either cases ($U=0$ eV). (c) The same dispersion map for a 1QL step with a $U= 0.8$ eV scalar potential overlapping with the step edge. In this case, the step edge state visible in panel (a) has been pushed out of the topological regime ($U>U_C$$\sim$0.3 eV) and a new topologically connected edge state indicated by the green arrow has emerged from the lower Dirac cone. (d) High symmetry points in the 1D step edge Brillouin zone are labeled $\overline{\Gamma}$ and $\overline{M}$, and overlaid on the 2D surface Brillouin zone. (e) Edge state velocity ($V_S$) at momenta slightly displaced from the 2D Dirac point ($K_\parallel \sim 0^+$) is plotted with circles as a function of the added step potential ($U$), and compared with dashed line fits from the 2D continuum-limit (CL) theory (Eq. \ref{eq-gapless-bound}). The step edge velocity is only well defined outside of the 2D state continuum ($-1<V_S/V_D<1$). The new edge state traced in green at at $E>0.4$ eV is indicated with a green arrow in panel (c). In the shaded regions, edge states connect from the 2D Dirac point to the (blue) lower or (orange) upper band structure continua.}
\label{fig:tight binding}
\end{figure}


To explore the phenomenology of the step edge state connectivity in a less idealized scenario, we briefly examine single- and double- quintuple layer step edges in a minimal 3D tight binding model. Bulk and surface state kinetics in the model are similar to vacuum-cleaved Bi$_2$Se$_3$, and it incorporates the same TI topological invariants ($[1;000]$) derived from a band inversion at the 3D $\Gamma$-point. In this approach popularized by Ref. \onlinecite{FuOriginal}, a single Bi$_2$Se$_3$ formula unit is reduced to two $p_z$ orbitals displaced along the surface normal axis. Parameters of the model are listed in Appendix \ref{appedix:TB}, and the projection of 2D momentum space onto the 1D momentum axis parallel to the step is shown in Fig. \ref{fig:tight binding}(d). The step aligns with the 2D $\Gamma$ - K axis, and creates a new periodicity from $\bar{\Gamma}$ to $\bar{\Gamma}$ which is traced in red in Fig. \ref{fig:tight binding}(d), with Kramers degeneracy required at the $\bar{M}$ point.

In this case, incorporating the physical one-quintuple-layer step drawn in Fig. \ref{fig:Structure}(a) involves modifying only the \emph{kinetic} Hamiltonian, but nonetheless results in the appearance of an edge state that connects between the Dirac point and the lower state continuum (see Fig. \ref{fig:tight binding}(a)). Unlike other effectively 1D topological in-gap states, the edge state shown in Fig. \ref{fig:Structure}(a) does not depend on extrapolation to a scenario with particle-hole symmetry, and actually vanishes when particle-hole symmetry if restored to the Hamiltonian (by setting $h_0=0$; see Appendix \ref{appedix:TB}). Increasing the height of the step to 2 quintuple layers results in a larger edge state group velocity near the Dirac point (Fig. \ref{fig:tight binding}(b)), but does not change the Dirac point or lower continuum connectivity. We note however that a sufficiently large step edge will effectively introduce a new 2D state continuum that overlaps with and obscures the 2D Dirac point.

Turning on a positive scalar potential ($U>0$) acting at nearest neighbor sites bordering the step causes the edge state group velocity to increase at the intersection point with the Dirac point (Fig. \ref{fig:tight binding}(e)), matching the behavior predicted in the Fig. \ref{fig:Step edge}(c) phase diagram. The edge state merges with the upper Dirac continuum above a critical potential of $U>U_C\sim 0.3 eV$, and ceases to connect to the Dirac point. However, rather than the system entering an extended `non-topological' phase region as posited in Fig. \ref{fig:Step edge}(c), a new Dirac-connected surface state emerges as an antibound state of the lower Dirac cone. This new state is indicated by a green arrow in Fig. \ref{fig:tight binding}(c), and is the Kramers partner of the original step edge state at the $\bar{M}$ point. Tracing the dispersion, we see that these two edge states effectively connect between the 2D Dirac point and the \emph{upper} band structure continuum, whereas the original state connected to the lower continuum. This topological distinction is indicated by shading in Fig. \ref{fig:tight binding}(e). 

Comparing with the analytic continuum-limit solution for a 2D Dirac surface with a delta function potential (CL model, see Appendix \ref{appendix:energy_cutoff}), we find that the analytic model gives a close match for the low-momentum dispersion as the step edge states emerge from the lower Dirac cone, but diverges as they approach the upper Dirac cone at $V_S/V_D>0.5$ in Fig. \ref{fig:tight binding}(e). For this plot, the analytic curves have been shifted to align with the tight binding model on the $U$-axis, and the input $U$ parameter has been rescaled upward by a factor of 3.9 for the left curve and 2.4 for the right curve. Taken together, this large upward rescaling of the $U$-axis, as well as the non-infinite critical potential for changing the Dirac point connection topology ($U_C\neq \infty)$, and the appearance of successive Dirac-connected states as a function of $U$, reveal that including 3D structure and coupling with bulk band symmetries can have an important role in defining the edge state properties. However, the rescaled 2D CL model can do an excellent job for edge states with group velocities not far removed from their band of origin (the lower or upper Dirac cone), and accurately describes the low-momentum dispersion for over half of the parameter space explored in Fig. \ref{fig:tight binding}(e).

\section{summary}\label{sec:summary}

In conclusion, we have established that step edge states in single-particle models of a 3D TI surface can manifest topological phase regions featuring protected connections to the 2D surface Dirac points, with no reliance on fine tuning in the bulk or surface Hamiltonian. This analysis builds on the previous observation of Dirac point connectivity in specific 2D models, and provides a guiding principle for understanding the likely form of in-gap states observed by STM at step edges. Realistically parametrized simulations of a Bi$_2$(Se/Te)$_3$ single-quintuple-layer step edge are performed using a 2D Dirac cone Hamiltonian and a 3D tight binding model, and establish that topologically connected step edge states are physically plausible in this material family.

\textbf{Acknowledgements:} We are grateful for discussions with F. Burnell, A. Kent, and P. Moon. This work was supported partially by the MRSEC Program of the National Science Foundation under Award Number DMR-1420073. R.R.B. was supported by Purdue University startup funds. L.A.W. acknowledges support from the NYU Global Seed Grant for Collaborative Research program. Materials development research at NYU is supported by NSF under MRI-1531664, and from the Gordon and Betty Moore Foundation's EPiQS Initiative through Grant GBMF4838. L.M. This research used resources of the Advanced Light Source, which is a DOE Office of Science User Facility under contract no. DE-AC02-05CH11231.

\appendix

\section{Exact continuum-limit model}\label{appendix:energy_cutoff}

In this section, we solve the exact continuum model for a 2D Dirac cone surface with a delta function potential, and a high energy cutoff. Imposing a high energy cutoff on the kinetic basis is shown to have no effect on bound state dispersion near the 2D Dirac point. Complementary derivations for very similar scenarios can be found in Ref. \onlinecite{Nagaosa_gate} and \onlinecite{JieLu_BS}, and the analytic real space wavefunction derivation in Ref. \onlinecite{JieLu_BS} can be manipulated to identify the localized $E^{-2}$ decay trend of the DOS projection onto high energy kinetic eigenstates. 

\subsection{The gapless Hamiltonian}

Consider a $(2+1)-D$ Dirac fermion in the presence of a singular delta potential parallel to the $y$-axis. In units where $\hbar$ and $v_{F}$ are unity, the effective Hamiltonian is
\begin{align}
\mc{H} = \mc{H}_{0} + \hat{V}, \; \mc{H}_{0} = \s_{x}\hat{p}_{x} + \s_{y}\hat{p}_{y}, \; \hat{V} = W \d(\hat{x}) .
\end{align}

Here, $W$ has units of energy times length, and is equivalent to the $U$ step potential parameter defined in Eq. \ref{eq:Udef} multiplied by the width of the potential barrier. Since translation symmetry exists along the $y$-axis, the eigenstates can still be labeled by the eigenvalues of $\hat{p}_{y}$. For a given eigenvalue $p_{y}$, we are thus left with a one-dimensional problem of a Dirac fermion in the presence of a delta function potential at the origin.

\subsection{The bound state energy: poles of the T-matrix}

\begin{figure}[h]
\begin{center}
\resizebox{10cm}{!}{\includegraphics{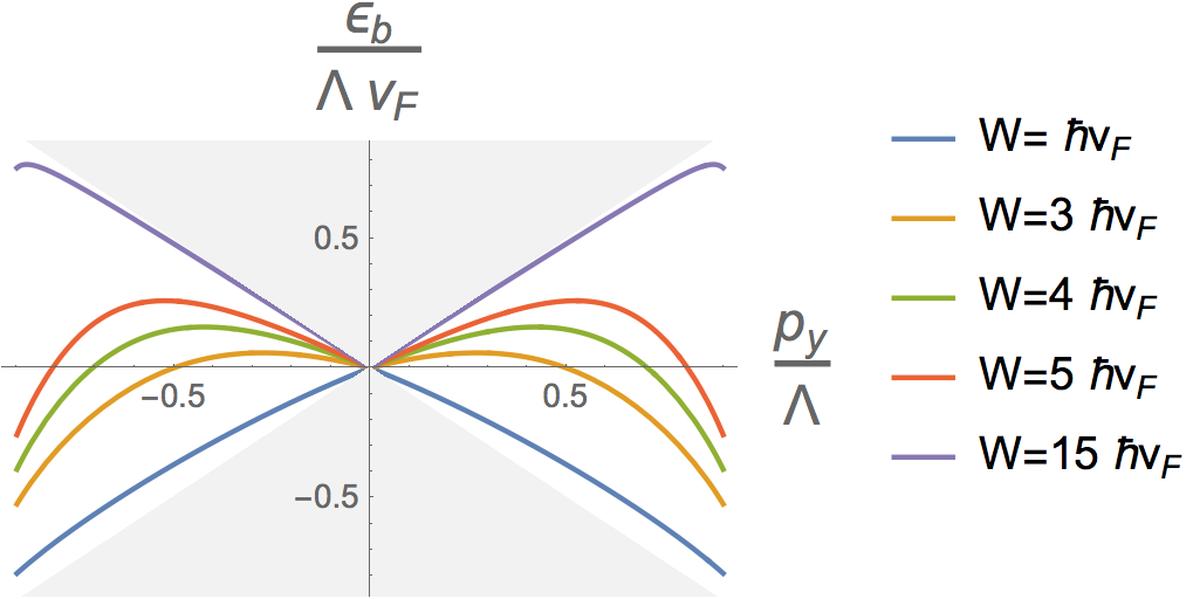}}
\caption{Bound state dispersions obtained by solving Eq.~\eqref{eq-gapless-bound}, with the substitution $\L \to \sqrt{\L^{2} - p_{y}^{2}}$. Units have been restored; $\L$ is the momentum cutoff, related to the energy cutoff, $E_{0}$, by the equation $E_{0} = v_{F}\L$. The bending effects are significant for intermediate values of $W \sim 4$. The 2D bulk bands occupy the shaded regions.}
\label{fig-enplot}
\end{center}
\end{figure}

The energy of the bound state, $\e_{b}(p_{y})$, can be obtained by finding the real poles of the $T$-matrix of the problem, defined via
\begin{align}
\hat{T}(E) = \le(1-\hat{V}\hat{G}_{0}(E)\ri)^{-1}\hat{V},
\end{align}
where $\hat{G}_{0}(E)$ is the retarded Green's function for the potential-free problem. For the Delta function potential, the $T$-matrix has a simple form $\le\la x\le|\hat{T}(E)\ri|x'\ri\ra = t(E) \d(x)\d(x')$, where
\begin{align}
t(E) = W\le(1- W \le\la 0\le|\hat{G}_{0}(E)\ri|0\ri\ra\ri)^{-1},
\end{align} 
and $\le\la 0\le|\hat{G}_{0}(E)\ri|0\ri\ra$ is the local on-site real space retarded Green's function,
\begin{align}
\le\la 0\le|\hat{G}_{0}(E)\ri|0\ri\ra = \int_{-\L}^{\L} \frac{dp_{x}}{2\p}\le(E - \le(\s_{x}p_{x} + \s_{y}p_{y}\ri) + i 0^{+}\ri)^{-1}.
\end{align}
The momentum cutoff, $\L$, has been introduced above. The analytic structure is simple when $E$ lies in the spectral gap, $E^{2} < (p_{x}^{2}+p_{y}^{2})$, where the bound state forms:
\begin{align}
\le\la 0\le|\hat{G}_{0}(E)\ri|0\ri\ra &= \int_{-\L}^{\L} \frac{dp_{x}}{2\p}\frac{E + \le(\s_{x}p_{x} + \s_{y}p_{y}\ri)}{E^{2} - \le(p_{x}^{2} + p_{y}^{2}\ri)} = \le(E + \s_{y}p_{y}\ri)\int_{-\L}^{\L} \frac{dp_{x}}{2\p}\frac{1}{\le(E^{2}-p_{y}^{2}\ri) - p_{x}^{2}}\nn\\
&= \le(E + \s_{y}p_{y}\ri)\frac{\arctan\le(\frac{\L}{\sqrt{p_{y}^{2} - E^{2}}}\ri)}{\p\sqrt{p_{y}^{2} - E^{2}}} \stackrel{\L\tto\infty}{=} \frac{1}{2}\frac{E + \s_{y}p_{y}}{\sqrt{p_{y}^{2} - E^{2}}}.
\end{align}
The bound state energy, given by the real poles of $t(E)$ within the spectral gap, is found via the condition
\begin{align}
\det \le(1- W \le\la 0\le|\hat{G}_{0}(\e_{b})\ri|0\ri\ra\ri) = 0, \text{ i.e., } \frac{\arctan\le(\frac{\L}{\sqrt{p_{y}^{2} - \e_{b}^{2}}}\ri)}{\p\sqrt{p_{y}^{2} - \e_{b}^{2}}} \le(\e_{b} \pm p_{y}\ri) = \frac{1}{W}.
\end{align}
The solution to this is given by
\begin{align}\label{eq-gapless-bound}
\e_{b} = \sgn(W)\le(\frac{w^{2} - 4}{w^{2} + 4}\ri) \le|p_{y}\ri|, \quad w = \frac{2|W|}{\p}\arctan\le(\frac{\L}{\sqrt{p_{y}^{2} - \e_{b}^{2}}}\ri) \stackrel{\L\tto\infty}{=} |W|.
\end{align}
This needs to be solved numerically to obtain the bound state dispersion at a finite $\L$. In order to maintain approximate rotational symmetry, $\L$ should be replaced by $\sqrt{\L^{2} - p_{y}^{2}}$. Some results are shown in Figure~\ref{fig-enplot}.

\subsection{The bound state dispersion: analytic results}

Taking $\L\to\infty$ and/or $p_{y}\to0$ yields a linear dispersion with a kink at the Dirac point,
\begin{align}
\e_{b} = v_{b} p_{y}, \quad v_{b} = \sgn(W p_{y})\le(\frac{W^{2} - 4}{W^{2} + 4}\ri).
\end{align}

Taking the first correction due to finiteness of $\L$, we have
\begin{align}
\e_{b} = \sgn(W)\le(\frac{W^{2} - 4}{W^{2} + 4}\ri) \le|p_{y}\ri| - \frac{128 W^{3}}{\p\L(W^{2} + 4)^{3}} p_{y}^{2} + \mc{O}(\L^{-2}).
\end{align}

\subsection{The massive Dirac Hamiltonian}

Consider the case when the Dirac fermion is massive, i.e.,
\begin{align}
\mc{H}_{0} = \s_{x}\hat{p}_{x} + \s_{y}\hat{p}_{y} + m \s_{z}.
\end{align}
In that case, when $E$ lies in the spectral gap, $E^{2} < (p_{x}^{2}+p_{y}^{2})$, where the bound state forms:
\begin{align}
\le\la 0\le|\hat{G}_{0}(E)\ri|0\ri\ra = \le(E + \s_{y}p_{y} + m \s_{z}\ri)\frac{\arctan\le(\frac{\L}{\sqrt{m^{2} + p_{y}^{2} - E^{2}}}\ri)}{\p\sqrt{m^{2} + p_{y}^{2} - E^{2}}} \stackrel{\L\tto\infty}{=} \frac{1}{2}\frac{E + \s_{y}p_{y}+ m \s_{z}}{\sqrt{m^{2} + p_{y}^{2} - E^{2}}}.
\end{align}

\begin{figure}[t]
\begin{center}
\resizebox{10cm}{!}{\includegraphics{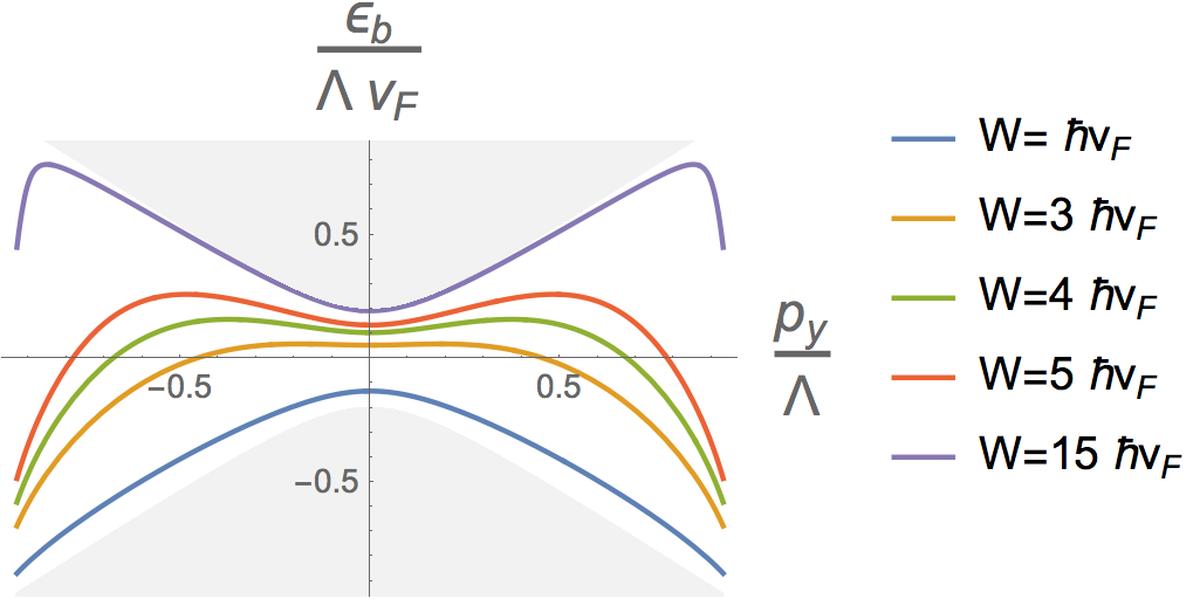}}
\caption{Bound state dispersions obtained by solving Eq.~\eqref{eq-gapped-bound}, with the substitution $\L \to \sqrt{\L^{2} - p_{y}^{2} - m^{2}}$ and $m=0.2 \L v_{F}$. Units have been restored; $\L$ is the momentum cutoff, related to the energy cutoff, $E_{0}$, by the equation $E_{0} = v_{F}\L$. The bending effects are significant for intermediate values of $W \sim 4$. The 2D bulk bands occupy the shaded regions.}
\label{fig-enplotG}
\end{center}
\end{figure}

Following the procedures outlined above for the gapless case, we obtain the bound state energy to be
\begin{align}\label{eq-gapped-bound}
\e_{b} = \sgn(W)\le(\frac{w^{2} - 4}{w^{2} + 4}\ri) \sqrt{p_{y}^{2} + m^{2}}, \quad w = \frac{2|W|}{\p}\arctan\le(\frac{\L}{\sqrt{p_{y}^{2} + m^{2} - \e_{b}^{2}}}\ri) \stackrel{\L\tto\infty}{=} |W|.
\end{align}
This needs to be solved numerically to obtain the bound state dispersion at a finite $\L$. In order to maintain approximate rotational symmetry, $\L$ should be replaced by $\sqrt{\L^{2} - p_{y}^{2}-m^{2}}$. Some results are shown in Figure~\ref{fig-enplotG}.

\subsection{The bound state dispersion: analytic results for the gapped case}

Taking $\L\to\infty$ yields a scaled relativistic dispersion:
\begin{align}
\e_{b} = \sgn(W)\le(\frac{W^{2} - 4}{W^{2} + 4}\ri)\sqrt{p_{y}^{2} + m^{2}}.
\end{align}

Taking the first correction due to finiteness of $\L$, we have
\begin{align}
\e_{b} = \sgn(W)\le(\frac{W^{2} - 4}{W^{2} + 4}\ri)\sqrt{p_{y}^{2} + m^{2}} - \frac{128 W^{3}}{\p\L(W^{2} + 4)^{3}} \le(p_{y}^{2} + m^{2}\ri) + \mc{O}(\L^{-2}).
\end{align}


\section{Three-dimensional tight binding model}\label{appedix:TB}

We adopt the minimal tight binding model framework for topological order, in which a single Bi$_2$(Se/Te)$_3$ formula unit within one quintuple layer of the crystal is reduced to two $p_z$ orbitals displaced along the surface normal axis \cite{FuOriginal,MaoPRB2011}. These two-orbital unit cells are arranged in an AA stacked hexagonal lattice, with an in-plane lattice constant of $a=4.2$ nm and a z-axis lattice constant of $c$, the precise value of which is not physically relevant.

\begin{figure}[t]
\includegraphics[width = 8.6cm]{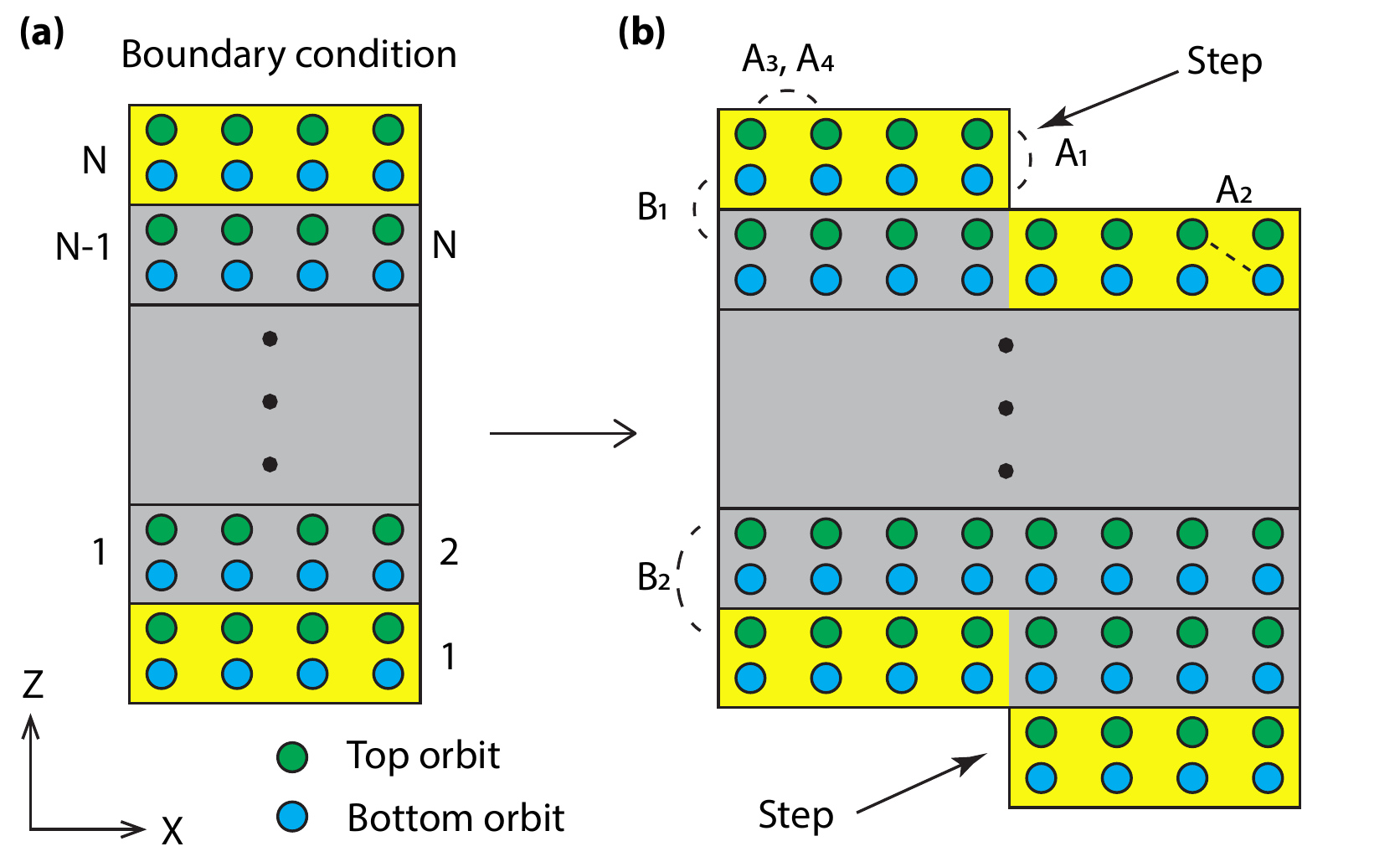}
\caption{{\bf{Tight binding orbital basis}}: (a) A cross section showing the $[\hat{z}, (\hat{x}/2+\hat{y}\sqrt(3)/2)]$ plane of a simulated slab containing $N_y=4$ in-plane sites. The upper and lower spin-degenerate orbitals in each unit cell shaded green and blue. Unterminated boundaries are labeled with numbers, indicating the slip-repeat connectivity of the system, and open in-plane boundaries representing 1 quintuple layer step edges are present in the upper right and lower left corners of the image. (b) A similar slab with $N_y=8$ in-plane sites and the step edge slip placed in the center. The real-space connectivity of hopping terms is indicated with dashed lines, and shading from panel (a) is preserved, for ease of comparison.}
\label{fig:SMTB}
\end{figure} 

Taking into account of the spin degree of freedom, the four states are $|P^{\pm}_z, \uparrow(\downarrow) \ra$, where $\pm$ and $\uparrow(\downarrow)$ indicate the parity and spin of the state respectively. To simulate a surface and step edge with translational symmetry along the x-axis, the modeling basis includes $N_z$=10 z-axis layers, which was sufficient to decouple the top and bottom surfaces. Along the $\hat{x}/2+\hat{y}\sqrt(3)/2)$ nearest-neighbor axis inside the plane of the surface, $N_y$=150 sites were used for full Brillouin zone spectral function maps, and a larger $N_y=1000$ system was used to eliminate finite size effects in all other panels. Coupling around the in-plane axis resulted in a small Dirac point gap that scaled approximately as $N_y^{-1}$, and had a value of 3 meV for $N_y=1000$. To avoid finite size effects in the analysis, the smallest nonzero momentum value considered in the manuscript has an amplitude of 0.005 $\AA^{-1}$, giving an intrinsic kinetic gap in the ideal 2D Dirac cone that is a factor of 6.7 ($6.7=\Delta_T/(3meV)$ larger than the finite size gap. The displayed plots at this momentum amplitude were qualitatively insensitive to $\sim50\%$ fractional changes in system size, which is understandable as for a gapped 2-state system, an off-diagonal matrix element with this relative amplitude would account for just a 1.1$\%$ change in the gap energy, and 0.55$\%$ mixing of PDOS.

A z-axis slip in the linkage of repeating boundary conditions is used to create the surface edge, as shown in Fig. \ref{fig:SMTB}. A Coulomb potential $U$ term was applied on the site closest to the step, to account for energetic factors that can not be described by the simple implementation of z-axis slip. This is described by a Hamiltonian term $H_U=U(\sum_i n_i)$, where $n_i$ is a number operator, and the sum over $i$ indexes all four orbital and spin states in the unit cell with an open in-plane boundary (i.e. the 2 real-space orbitals connected by the $A_1$ dashed line in Fig. \ref{fig:SMTB}(b)). Though the top and bottom surfaces were effectively decoupled, an identical $H_U$ term was applied to the step edge on the bottom of the slab for the sake of symmetry. When represented in real space, the hopping Hamiltonian term around the boundary includes a phase factor proportional to $k_\parallel$, due to the fact that the plane of the slab is not orthogonal to the step axis (x-axis).

In this model, there are two types of hopping -- intra-layer and inter-layer hopping terms of which the vectors connecting the corresponding unit cells are represented by $\vec{a}_{1,2,3}$ and $\vec{a}_{4}$:
\begin{align*}
    \vec{a}_1&=(a,0,0),  & \vec{a}_2&=(\frac{a}{2},\frac{\sqrt{3}a}{2},0), \\
    \vec{a}_3&=(-\frac{a}{2},\frac{\sqrt{3}a}{2},0),
     & \vec{a}_4&=(0,0,c).
\end{align*}

where $a$ and $c$ are the in-plane and out-of-plane lattice parameters respectively.

The Hamiltonian for the tight-binding model can be written as
\begin{equation}
H= \sum_{i}   \tilde{\epsilon}_i  + (\sum_{\langle i < j\rangle} \tilde{t}_{ij} + h.c.)
\end{equation}
where $\tilde{\epsilon}$ describes the on-site energetics, $i,j$ index nearest neighbor unit cells. The hopping matrix $\tilde{t}_{ij}$ consists of elements
\begin{equation*}
 \le\la \vec{r}_{i}, m'_{\tau}, m'_{\sigma}\le| \tilde{t}_{ij} \ri|\vec{r}_{j}, m_{\tau}, m_{\sigma}\ri\ra,
\end{equation*}
where $\vec{r}_{i}, \vec{r}_{j}$ indicates the lattice vectors, $m_{\tau}$ indexes the upper and lower orbitals (blue vs. green in Fig. \ref{fig:SMTB}), and $m_{\sigma}$ indicates spin up or spin down, quantized on the z-axis. 

If 3D translational symmetry is assumed, the tight-binding Hamiltonian takes the following form in momentum space
\begin{align}
H & = \sum_{\vec{k}} H(\vec{k}) \\
H(\vec{k}) & = \tilde{\epsilon} + (\sum_{i=1}^{4} \tilde{t}_{\vec{r}_i,\vec{r}_i+\vec{a}_{i}} e^{i \vec{k}\cdot \vec{a}_{i}} + h.c.)
\end{align}

which can be presented as

\begin{equation}\label{eq:tb}
    H(\vec{k}) = h_{0} + h_{1} \Gamma_{1} + h_{2} \Gamma_{2}+h_{3} \Gamma_{3}+h_{4} \Gamma_{4}
\end{equation}
where $\Gamma_{i}$ is given as an outer product of two sets of Pauli matrices $\tau_{i}$ and $\sigma_{i}$ by
\begin{align*}
 \Gamma_{1} & = \tau_{x} \otimes \mathbf{1}, & \Gamma_{2} & = \tau_{y} \otimes \mathbf{1},\\
 \Gamma_{3} & = \tau_{z} \otimes \sigma_{x},& \Gamma_{4} & = \tau_{z} \otimes \sigma_{y}.
\end{align*}

The coefficients $h_{i}$ in Eq. (\ref{eq:tb}) are as follows
\begin{align*}
    h_{0}  = & 2A_{4}\big[\sum_{i=1}^{3} \cos(\vec{k} \cdot \vec{a}_{i})\big] + 2 B_{2} \cos( \vec{k} \cdot \vec{a}_{4}) \\
    h_{1} = &2 A_{1} + 2 A_{2} \big[\sum_{i=1}^{3} \cos(\vec{k} \cdot \vec{a}_{i})\big] + \\
    & B_{1} \cos( \vec{k} \cdot \vec{a}_{4}) \\
    h_{2} = & B_{1} \sin( \vec{k} \cdot \vec{a}_{4}) \\
    h_{3}  = &    -\sqrt{3} A_{3} \big[\sin(\vec{k} \cdot \vec{a}_{2}) + \sin(\vec{k} \cdot \vec{a}_{3})\big] \\
    h_{4} = &A_{3} \big[ 2 \sin(\vec{k} \cdot \vec{a}_{1}) + \sin(\vec{k} \cdot \vec{a}_{2}) -\sin(\vec{k} \cdot \vec{a}_{3}) \big] 
\end{align*}

where $A_{i}$ and $B_{i}$ are the parameters for the intra-layer and inter-layer hopping terms respectively. In our simulation, the parameters are given as following:
\begin{align*}
     A_{1} & = -0.41, & A_{2} & = 0.17,& A_{3} & = 0.15,  \\
    A_{4} & = -0.055, & B_{1} &= 0.4, & B_{2} &= 0.05.
\end{align*}

With the above hopping parameters, the resulting model is a strong 3D TI with topological invariants $[1;000]$ and band inversion at the 3D $\Gamma$ point as for Bi$_2$Se$_3$-family TIs. The bulk band gap is 0.3 eV, similar to that of Bi$_2$Se$_3$, and the Dirac cone has a velocity of $\sim 2$ eV$ \cdot \AA^{-1}$.

\bibliography{TI_step_references}
\bibliographystyle{apsrev4-1}

\end{document}